\documentclass[superscriptaddress,amssymb,twocolumn,aps]{revtex4}

\usepackage{graphicx,dcolumn,bm,amsmath,amssymb,color,bm}

\usepackage[utf8]{inputenc}
\usepackage[T1]{fontenc}
\usepackage{mathptmx}
\usepackage{gensymb}

\usepackage{hyperref}
\usepackage{color}

\newcommand{\eq}[1]{Eq.~(\ref{#1})}

\newcommand{\fig}[1]{Fig.~\ref{#1}}

\newcommand{\red}[1]{ {\color{black} #1}}

\definecolor{amber}{rgb}{1.0, 0.75, 0.0}

\newcommand{\eeq}{ \end{equation} }
\newcommand{\beq}{ \begin{equation} }

\newcommand{\eea}{ \end{align} }
\newcommand{\bea}{ \begin{align} }

\newcommand{\oma}{\vec{s}}

\newcommand{\bor}{ {\bf r} }

\usepackage{xr}
\makeatletter
\newcommand*{\addFileDependency}[1]{
  \typeout{(#1)}
  \@addtofilelist{#1}
  \IfFileExists{#1}{}{\typeout{No file #1.}}
}
\makeatother
\newcommand*{\myexternaldocument}[1]{%
    \externaldocument{#1}%
    \addFileDependency{#1.tex}%
    \addFileDependency{#1.aux}%
}
\myexternaldocument{supplementary}

\begin{document}

\preprint{AIP/123-QED}


\title{Freezing in flat monolayers of soft spherocylinders}

\author{Jaydeep Mandal}
\affiliation{%
Centre for Condensed Matter Theory, Department of Physics, Indian Institute of Science, Bengaluru 560012, India
}%

\author{Henricus H. Wensink}
\affiliation{Laboratoire de Physique des Solides - UMR 8502,
CNRS, Universit\'e Paris-Saclay, 91405 Orsay, France}

\author{Prabal K. Maiti}
\email{maiti@iisc.ac.in}
\affiliation{%
Centre for Condensed Matter Theory, Department of Physics, Indian Institute of Science, Bengaluru 560012, India
}%
\date{\today}

\begin{abstract}

 Lamellar or smectic phases often have an intricate intralamellar structure that remains scarcely understood from a microscopic viewpoint.  In this work, we use molecular dynamics simulations to study the effect of volume exclusion and electrostatic repulsion on the phase transitions of a flat membrane of soft  spherocylinders. With increasing rod packing, we identify nematic and solid phases and find that the nematic-crystal phase transition happens at a uniform packing fraction (\(\eta_c \approx 0.82\)), independent of the spherocylinder aspect ratio. \red{This value is considerably higher than the well-known critical freezing transition of a hard disk fluid ($\eta_{c} \approx 0.7$) to which one could naively map a system of near-parallel rods with co-planar mass centers. We attribute this difference to a non-vanishing residual orientational entropy per rod.} Our findings are corroborated by a simple theory based on a simple microscopic density functional theory of freezing of a two-dimensional rod fluid.  Introduction of electrostatic interactions between the rods reduces the lateral compressibility of the monolayer fluid but keeps the positional order unhindered, which in turn maintains the packing fraction at the nematic-crystal transition. The strength of the orientational fluctuations of the individual rods in our membranes exhibits a density scaling that differs from  3D bulk smectics. Our findings contribute to a qualitative understanding of liquid crystal phase stability in strong planar confinement and engage with recent experimental explorations involving nanorods on 2D substrates.
 
\end{abstract}

\maketitle


\section{\label{sec:intro}Introduction:}

Liquid crystals are states of matter exhibiting intermediate symmetries between the crystal and liquid phases \cite{de1993physics}. They exhibit broken symmetry and long-ranged ordering in orientational phase space, such as in a nematic state, and on top of that may have long-ranged positional order in periodicity in one dimension such as a smectic phase, or in two such as in a columnar. Apart from being fascinatingly complex materials to study in their own right, the coexistence of order and fluidity endows these materials with unique, often reconfigurable optical and mechanical properties that have a strong potential for use in various novel soft matter applications going beyond well-established optical display technology \cite{woltman2007liquid,lagerwall2012new}.

Microscopic models aim at elucidating the various phase transformation and microscostructure in relation to the mesogen shape \cite{lansac2003phase, maiti2002induced,maiti2004entropy}. The simplest one is the uniaxial, cylindrical rod  which has inspired Onsager in his seminal work on the entropy-driven isotropic-nematic phase transition \cite{onsager1949effects}. Decades later, the advent of computer simulations has enabled a much broader systematic survey of LC phases \cite{maiti2009computer,bag2016atomistic,naskar2020liquid,saurabh2017understanding,chattopadhyay2021heating,earl2001computer,cuetos2002monte} and their phase transitions including smectic and crystals phases in relation to the length-to-width ratio \cite{frenkel1988thermodynamic,bolhuis_frenkel,mcgrother1996re,lopes2021phase}. Similar numerical explorations have been undertaken for disc-shaped mesogens \cite{veerman1992phase,duncan2009theory,marechal2011phase,bag2015molecular} where columnar phases are prevalent at high pressures. The common thread among these models is that mesogens interactions are taken as strictly hard in which case all order-disorder transition are driven by entropy alone \cite{lekkerkerker1998ordering,frenkel2015order,chattopadhyay2023entropy}. Since temperature plays no role at all, the (effective) shape anisotropy in combination with the packing fraction of the mesogens are the key parameters in determining the phase diagram.   



Among the various LC states, smectic (Sm) or lamellar phases  are of particular interest \cite{de1993physics} since the particles are arranged in layers, with the long molecular axes approximately perpendicular to the laminar planes. The only long-range order extends along this axis; with the result that individual layers can slip over each other (hence the "soap-like" nature) in a manner similar to that observed in graphite. The unidimensional periodicity imparts important photonic characteristics as it is tunable by varying the mesogens concentration which controls the smectic layer spacing. In case of colloidal sized rods or hyperswollen lamellar phases \cite{el2021fine} the typical spacing is of the order of the wavelength of visible light \cite{yamamoto2005dynamic}.

Another level of complexity arises from the way the rods are organized within the individual layers. The simplest cases are the SmA \cite{halperin1974first} where rods adopt a planar fluid arrangement and co-align with the layer normal,  or a SmC \cite{cabib1977smectic,leslie1991continuum,mishra2024glassy} where rods are tilted with respect to the normal. A state of lower symmetry can be realized if the rods adopt some crystalline in-plane lattice such as hexagonal in case of a SmB material. Many other smectic flavors exist featuring combinations of quasi-long-ranged or long-ranged in-plane ordering, director tilt and chirality which we will not discuss further here \cite{de1993physics}.

Despite a wealth of experimental characterizations mostly in the thermotropic domain of low-molecular weight mesogens \cite{lagerwall2006current}, microscopic models capable of reproducing the complex in-plane ordering of smectic LCs remain scarce \cite{meyer1974simple,de1972tentative,coldwell1974free}. Recently, several smectic phases with complex in-plane order have been identified in molecular simulations of polymer-grafted colloids \cite{fall2024complex}.  In this paper we wish to address a benchmark case by studying an individual smectic layer composed of freely rotating spherocylinders and explore  phase transitions the rods undergo upon increased in-plane crowding.  Knowledge of disorder-order transitions in these monolayers will not only be helpful to further develop microscopic theoretical theories of smectic phases \cite{wensink2023elastic}, it also engages with recent experimental and modelling advances  on colloids on substrates where anisotropic colloids with a variety of shapes can be experimentally realised \cite{sacanna2013engineering,qi2012ordered,florea2014towards,pujala2015spontaneous,herod1998two,kim2001langmuir,davies2014assembling,oettel2016monolayers,klopotek2017monolayers}.

In this work, we use Molecular Dynamics (MD) simulation supplemented with a simple density functional theory (DFT) to study the behaviour of a system of soft repulsive spherocylinders (SRS, often termed simply as "rods" in the manuscript), the center of masses (COMs) of which are confined to a planar surface (see \fig{fig:fig0_schematics}).  As such the system represents the limiting case of spherical monolayers recently explored in Ref \cite{rajendra2023packing} where the shell radius is taken to be infinitely large. We demonstrate that at non-vanishing planar rod concentrations, such monolayer fluids only exhibit two principal phases, namely a nematic (N) liquid crystal and a solid (K) phase.
The main questions we wish to address in our work are; what is the critical in-plane packing fraction for in-plane freezing? 

How does the onset of freezing change with the rod aspect ratios and strength of orientational fluctuations? and how does the critical packing fraction compare with values established for other benchmark models for entropy-stabilized liquid crystals? Further emphasis shall be put on the microstructure  and orientational probability of the individual rods across the range of packing fractions.  We also wish to explore the effect of electrostatic twist relevant in many colloidal rod systems where the colloids almost invariably carry surface charges that can sometimes be effectively subsumed into an effective hard rod description \cite{grelet2014hard}.


The rest of the manuscript is organized as follows. We begin in Section \ref{sec:simulation_details} by  describing the simulation model.  The main results for monolayer systems of neutral rods are presented in section Section \ref{sec:results} in which we discuss the freezing transition and the orientational order of the rods within the monolayer. In Section \ref{sec:electro} we assess the role of electrostatic repulsion on these properties. In the next Section \ref{sec:dft} we propose a simple density functional theory to predict the onset of freezing independent from the rod aspect ratios.   Finally, we summarise our findings and conclude in Section \ref{sec:conclusion}.

\section{Simulation Details} \label{sec:simulation_details}

\begin{figure}[]
    \centering
    \includegraphics[width=0.8\linewidth]{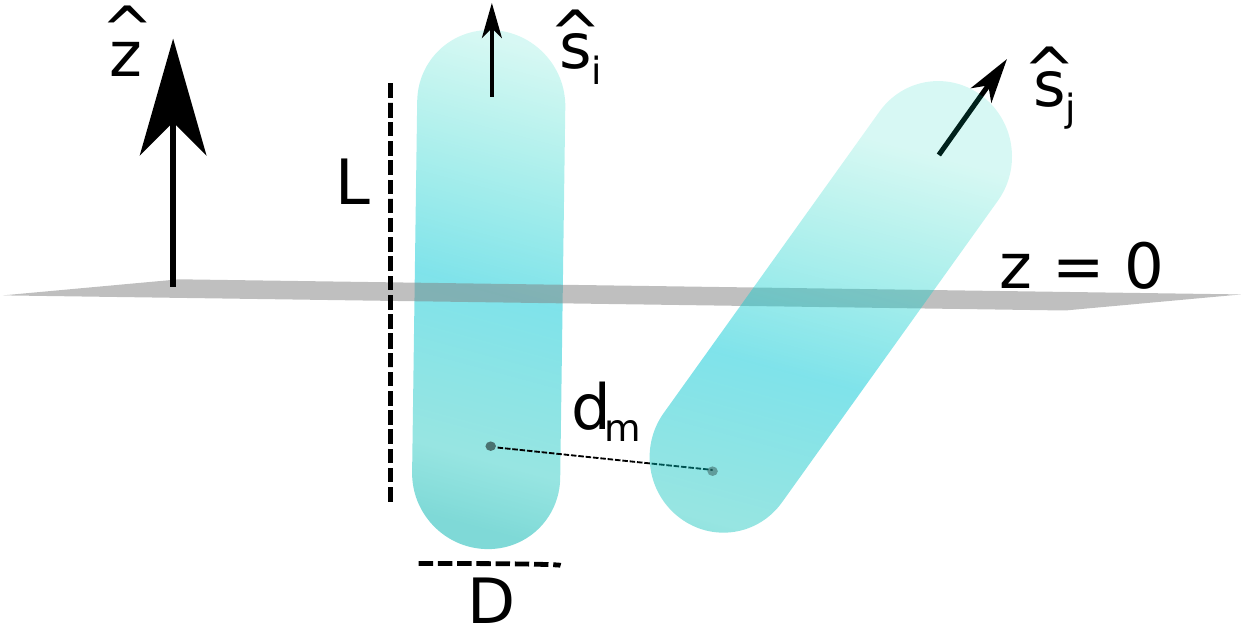}
    \caption{Schematic description of monolayer fluid of freely rotating rods represented by spherocylinders whose center of mass are constrained on the surface of the \(z=0\) plane. Shown are the orientation vectors $\hat{s_i}$ and $\hat{s_j}$ of the spherocylinders and the distance of closest approach \(d_m\) between the particle surfaces. }
    \label{fig:fig0_schematics}
\end{figure}

\begin{figure*}[]
    \centering
    \includegraphics[width=1.0\linewidth]{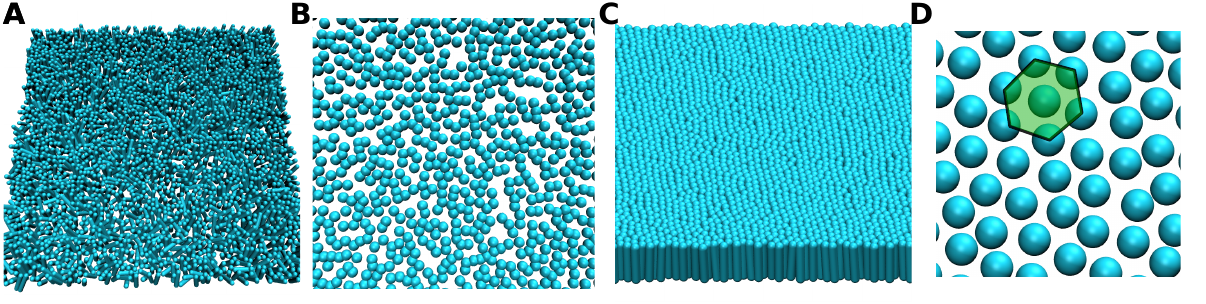}
    \caption{{Equilibrium configurations of a rod monolayer   with \(N = 2500, L/D = 5.0\)}. (A) Nematic fluid at packing fraction $\eta = 0.3$.  (B) Top view of the rod mass centers. (C)  Crystalline solid at \(\eta=0.9\)  featuring both positional and orientational ordering. (D) Hexagonal arrangement of the rod mass centers within the solid. For clarity, a hexagonal arrangement is shown by a green-shaded area. Visualizations are obtained using VMD software \cite{humphrey1996vmd}.   
    }
    \label{fig:Fig1_initial}
\end{figure*}

In this work, we consider a system of soft repulsive spherocylinders (SRSs) with unrestricted orientational freedom, the COMs of which are constrained on the surface of a perfectly flat two-dimensional plane (see \fig{fig:fig0_schematics}). Spherocylinders are geometric objects consisting of cylinders of length \(L\) and diameter \(D\), whose ends are capped with a hemisphere of the same diameter. Therefore, the total length of the spherocylinder is \(L+D\), but the length of the axis is \(L\). The aspect ratio of the spherocylinders is defined as \(A = L/D\). The spherocylinder mass is uniformly distributed and hence the center of mass of the particles lies at a distance $(L+D)/2$ from the tip of the spherocylinder. 
The two-dimensional center of mass of the \(i\)-th particle is denoted by \(\vec{r_i}\), and the orientation of the particle is denoted by \(\vec{s_i}\). The equation of motion for the variables are as follows,
\begin{equation} \label{eqn:pos_eom}
    \frac{d\vec{r_i}}{dt} = \vec{v_i}, \quad
    m\frac{d\vec{v_i}}{dt} = -\nabla_i U_{tot} 
\end{equation}
\begin{equation} \label{eqn:ornt_eom}
    \frac{d\vec{s_i}}{dt} = \vec{\omega_i} \times \vec{s_i}, \quad
    I\frac{d\vec{\omega_i}}{dt} = \vec{\tau_i}, 
\end{equation}
where \(U_{tot}\) is the total interaction energy determined by its interaction  with all surrounding particles,  \(\vec{v_i}\) and \(\vec{\omega_i}\) respectively denote its translational and rotational velocity, \(I\) the moment of inertia tensor and \(\tau_{i}\) is the torque experienced by particle $i$. The center of masses of the particles always obey the constraint  \(z_i = 0\) which enforces the rod COMs to remain co-planar. 


The interactions between the spherocylinders are represented by a  steeply repulsive WCA pair potential \cite{weeks1971role},
\begin{align}
    U_{\rm WCA}(d_{m}) = \begin{cases}
            4\varepsilon\left[ \left( \frac{D}{d_m} \right)^{12} - \left( \frac{D}{d_m} \right)^6 \right] +\varepsilon, &d_m < 2^{1/6} D \\
            0 , \hspace{1cm} d_m\geq 2^{1/6}D
        \end{cases}
        \label{wca}
\end{align}
Here, $d_{m}$ denotes the  distance of closest approach \cite{vega1994fast} between the spherocylinders (\fig{fig:fig0_schematics})), and $\epsilon$ is the associated energy.
In order to arrive at a more realistic description of colloidal nanorods where surface charges are often present we complement the reference WCA part with an effective line charge potential that results in a softer potential \cite{stroobants1986effect},
\begin{equation} \label{eqn:eqn1_el}
    U_{el}(\gamma, d_{m}) = u\frac{e^{-\kappa d_m}}{|\sin \gamma|}, \hspace{0.3cm} d_m\geq 2^{1/6}D,
\end{equation}
where $u >0$ denotes the amplitude of the electrostatic interaction, $\kappa^{-1}$  the Debye screening length and $\gamma$ is the relative angle between a rod $i$ and $j$ ($\cos \gamma = \oma_{i} \cdot \oma_{j}$). The amplitude $u$ can be related to a number of characteristics such as the effective line charge density on the rods, and basic length scales such the Bjerrum length, rod diameter and Debye screening length. Explicit expression for $u$ are discussed in Ref. \cite{stroobants1986effect} but in our model we will keep as it as a simple amplitude quantifying the extent of electrostatic repulsion.   The key impact of the potential, however,  relates to the inverse $|\sin \gamma |$ which diverges for strictly parallel rod configurations. This  embodies  the so-called {\em electrostatic twist} effect which penalizes near parallel pair configuration in view of the significant overlap of double layers involved.  Strictly, the above potential only holds for infinitely large aspect ratio rods and corrections must be anticipated for shorter rods \cite{eggen2009effective}.

We assume pair-wise 
additivity and define the total potential energy as a sum of the pair contributions via $U_{tot} = \tfrac{1}{2} \sum\sum_{j \neq i} (U_{\rm WCA}^{(ij)} +  U_{el}^{(ij)} )$ and use standard Molecular Dynamics (MD) simulations \cite{frenkel2023understanding} with \(N = 2500\) particles. We  employ an isothermal-isobaric (NPT) ensemble at fixed  number of particles \(N\), pressure \(P\) and temperature \(T\), using a Berendsen thermostat and barostat to maintain temperature and pressure \cite{berendsen1984bath}. A velocity-Verlet algorithm \cite{swope1982computer} for linear molecules \cite{rapaport2004art} is used for numerical integration, with an adaptation of the RATTLE algorithm \cite{andersen1983rattle,paquay2016method}, to maintain the planar confinement. The timestep is fixed at  \(\delta t = 0.001\) and  \(5 \cdot 10^5\) steps are used for both equilibration and production runs for each state point.  The relaxation time for mainiting temperature and pressure were respectively \(\tau_T = 50 \cdot \delta t\) and \(\tau_P = 1000\cdot \delta t\).

All quantities, thermodynamic and structural, are scaled by the elementary energy and length-scales of the system, $\varepsilon$, $D$ and $L$, and expressed in reduced units: temperature $T^* = k_BT/\varepsilon$ with $k_{B} T$ the thermal energy in terms of Boltzmann's constant $k_{B}$, pressure $P^* = aP/(k_BT)$, packing fraction $\eta = a\rho$, where $\rho = N/{\mathcal A}$ is the rod density with ${\mathcal A}$ the monolayer area and $a = \pi D^2/4$  the cross-sectional area of the spherocylinder. We measure time in units of $D(m/\varepsilon)^{1/2}$. In our calculations, we take the value of the Boltzmann constant \(k_B = 1.0\), and interaction parameters \(\epsilon,\sigma = 1.0\), and also the system parameters, \(D,m = 1.0\) for computational convenience.

In order to measure the orientational order of the rods we monitor the traceless symmetric tensor $Q$ defined as \cite{de1993physics},
\begin{equation} \label{eqn:nematic-order-parameter}
    Q_{\alpha \beta} = \frac{1}{N}\sum\limits_{i=1}^{N} \frac{3}{2} s_{i \alpha} s_{i \beta} \ -\ \frac{1}{2}\delta_{\alpha \beta},
\end{equation}
where $i$ is the particle index while $\alpha,\beta$ corresponds to components of unit orientation vector $\vec{s}$. 
The scalar nematic order parameter $S$ is the largest eigenvalue of $Q$, and its corresponding (three-dimensional) eigenvector $\vec{n}$ denotes the director of the ordered phase. As in conventional liquid crystals \cite{de1993physics} we may distinguish isotropic order having values of $S$ close to zero from states with strong orientational order where $S$ will be  close to unity. 

\section{Nematic-solid transition for neutral rods } \label{sec:results}


We begin by considering neutral SRSs ($u=0$) in which case rod interactions are driven primarily by volume-exclusion effects.  The system of a flat monolayer then exhibits only two phases at increasing packing fractions (\fig{fig:Fig1_initial}): a) The nematic phase at low to medium packing fractions (\(\eta \lessapprox 0.82\)) characterized by long-ranged orientational ordering but with no positional ordering  and b) the crystalline phase at high packing fractions where the particles are both orientationally and positionally ordered. In the nematic phase,  the mass centers adopt fluid structure within the monolayer plane (\fig{fig:Fig1_initial}B), combined with weak to moderate orientation order (\fig{fig:Fig1_initial}A). In the crystal phase, the mass centers are located at hexagonal lattice on the planar surface (\fig{fig:Fig1_initial}D), and oriented along the normal to the surface (\fig{fig:Fig1_initial} C). These two phases are observed for all SRS aspect ratios considered.


\begin{figure}[]
    \centering
    \includegraphics[width=1.0\linewidth]{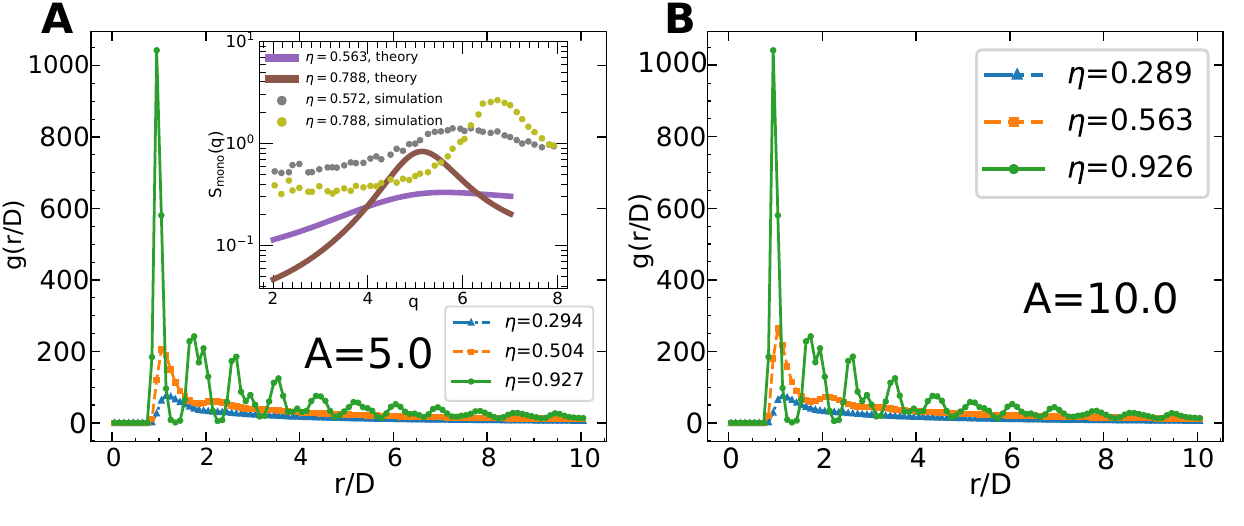}
    \caption{ Radial distribution function (RDF) of the rod mass centers for a rod monolayer. Shown are results for  two different aspect ratios \(L/D = 5.0\) (A) and \(10.0\) (B) each at three different packing fractions. At high packing fractions \red{crystalline  order emerges} whereas the liquid structure is observed at low to medium packing fractions. \red{The inset in fig A depicts the structure factor $S_{\rm mono}(q)$ for a nematic fluid of rods with $A=5$ at various packing fractions. Symbols denote simulations, solid lines the density functional theory (Section V).}
    }
    \label{fig:Fig2_gofr}
\end{figure}

\begin{figure}[]
    \centering
    \includegraphics[width=0.8\linewidth]{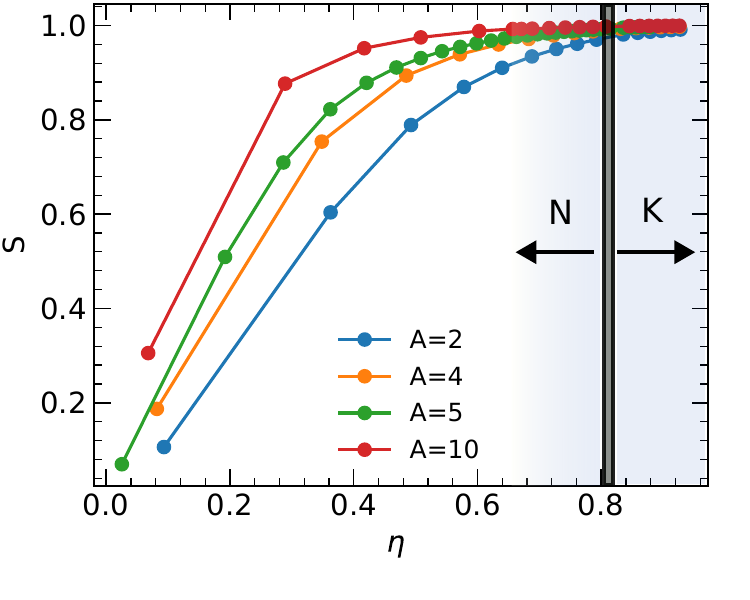}
    \caption{Nematic order parameter \(S\) versus packing fraction (\(\eta\)) for a monolayer of neutral rods with various aspect ratio (\(A\)). For a specific \(\eta\), the values of order parameter is larger for larger \(A\), indicating better ordering for longer rods. A smooth increase is observed for all rod shapes. The nematic (N) to solid (K) transition is indicated by a black solid line. 
    }
    \label{fig:Fig3_nemop}
\end{figure}

We may quantify the different phases with the help of standard correlation functions for the rod mass centers. In \fig{fig:Fig2_gofr} we show the radial distribution function (RDF) for the COM-s of the particles. \fig{fig:Fig2_gofr}(A) and \ref{fig:Fig2_gofr}(B) shows the RDF at three different packing fractions for the aspect ratio \(A = 5.0\) and \(A=10.0\) respectively. At low packing fractions, the radial distribution function exhibits fluid-like behavior for the centers of mass of the particles on the plane. The lack of positional ordering suggests a positionally disordered fluid. However, at high packing fractions, multiple peaks appear, indicating the formation of an ordered crystals of the particle centers of mass.  

 The orientational ordering can be quantified using the nematic order parameter \(S\). In \fig{fig:Fig3_nemop}, we demonstrate the behaviour of \(S\) with packing fraction for the different aspect ratio values of the spherocylinders. We observe that at low packing, the value of \(S\) is small, but rapidly increases as packing increases. Larger values of \(S\) suggest that rods are strongly aligned along the plane normal. 
 As expected, orientational order is stronger for more elongated rods at any given packing fraction whereas $S$ tends to saturate to near unity for all values of $A$ near close packing.
 
 \fig{fig:Fig3_nemop} also suggests that a stable isotropic phase ($S=0$) seems only possible at infinite dilution. We do not observe the scaling relation \(S \propto 1/\sqrt{N}\) at low densities, where \(N\) is the measure of system size. 
 We argue that the reason for the absence of any isotropic phase (characterized by an absence of  orientational or positional ordering at large length scales) relates to the entropy of the system. To maximize the packing entropy, the free surface fraction \(\phi_p\) (projection of rods on the plane) should be minimized, which can be achieved by allowing the particles a small but finite amount of orientational order by favouring certain normal alignments at any packing fraction except for infinite dilution where rod-rod correlations are virtually absent. The isotropic phase is thus expected to be stable only at extremely low packing fractions but remains elusive\cite{varga2016effect,martinez2014phase} for the finite-density systems we consider. Also, the low packing fraction regime is of little practical interest for most cases (such as lamellar phases where intralayer crowding is strong)  and  we will not discuss this issue any further in this work.


\begin{figure}[]
    \centering
    \includegraphics[width=1.0\linewidth]{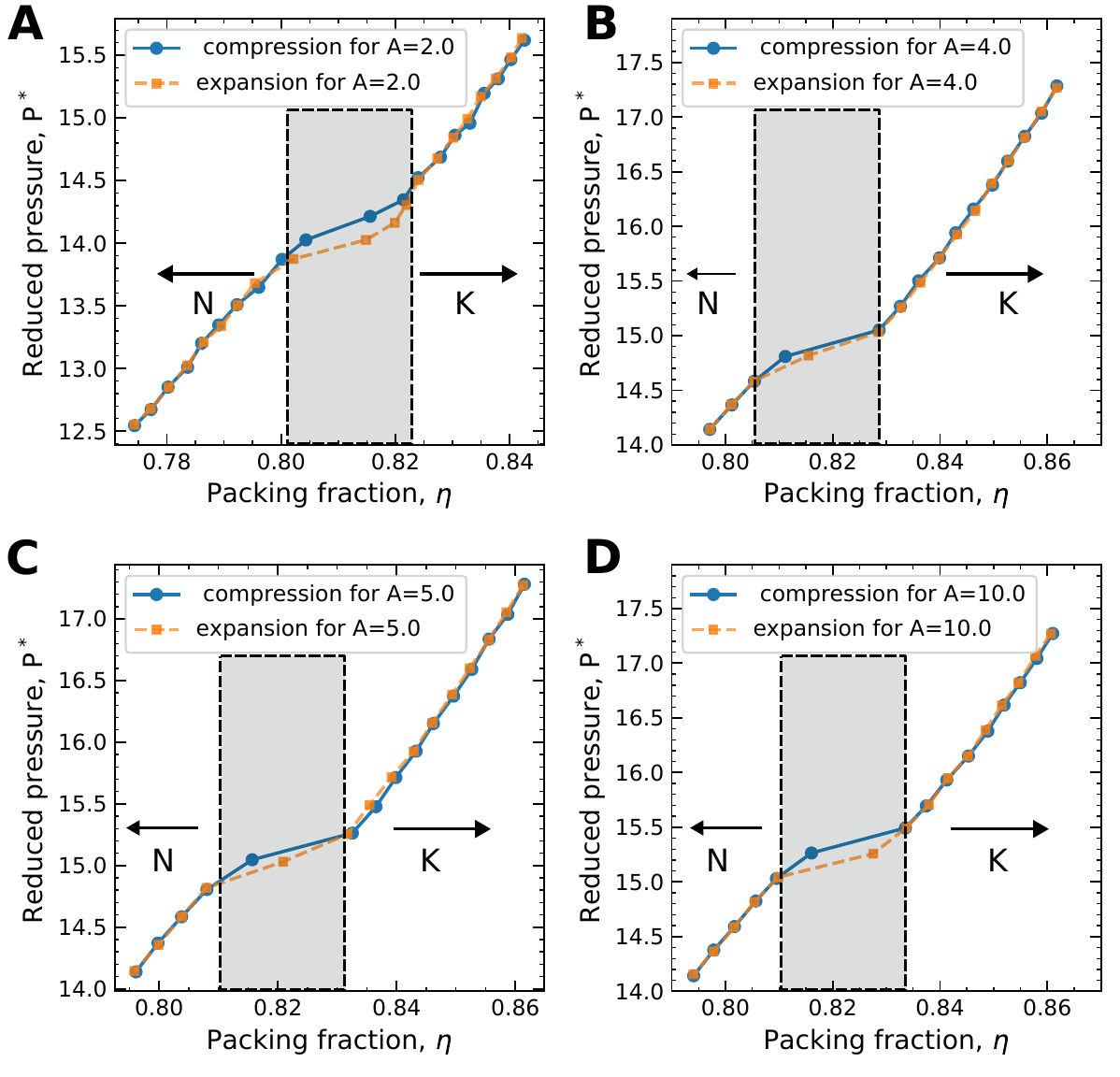}
    \caption{ Equation of state for monolayer rods for four different aspect ratios; \(A = 2.0\) (A),  \(A = 4.0\) (B),  \(A = 5.0\)  (C) and \(A = 10.0\) (D). A hysteresis signalling nematic to solid freezing transition is observed at around a packing fraction of \(\eta_c \approx 0.82\) irrespective of the aspect ratios. The nematic (N) to solid (K) transition region is indicated by the shaded area.
    }
    \label{fig:fig3_eos}
\end{figure}

\begin{table}[]
    \centering
\begin{tabular}{ |c | c | } 
 \hline
 \textbf{\(A=L/D\)} & \textbf{\(\eta_c\)}   \\ \hline
 2.0    & 0.81 $\pm$ 0.01  \\ \hline 
 4.0    & 0.82 $\pm$ 0.01  \\  \hline
 5.0    & 0.82 $\pm$ 0.01 \\ \hline
 10.0    & 0.82 $\pm$ 0.01 \\ \hline
    \end{tabular}
    \caption{Packing fraction \(\eta_c\) for the freezing transition for different aspect ratios \(L/D\) obtained from  \fig{fig:fig3_eos}. }
    \label{tab:table_transitionPF}
\end{table}

\begin{figure*}[]
    \centering
    \includegraphics[width=0.95\linewidth]{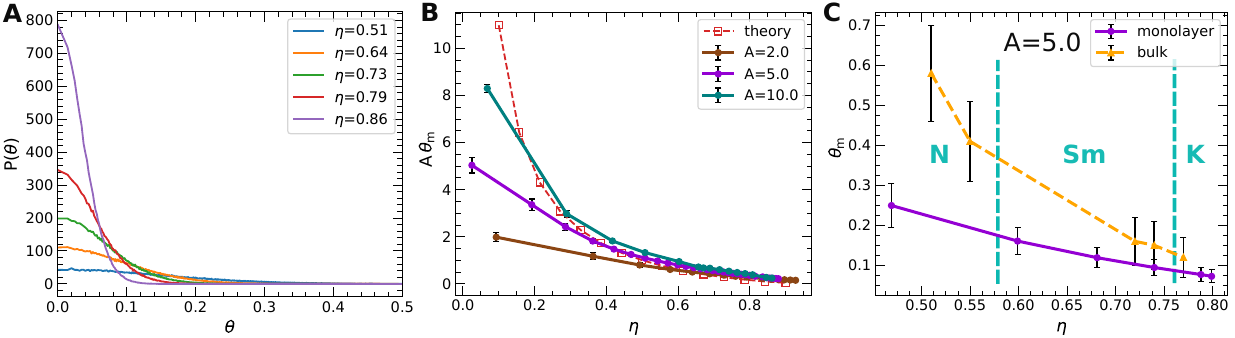}
    \caption{A) Normalized distributions of the polar angle around the layer normal for the aspect ratio \(A = 5.0\). The distribution for \(A=2.0,10.0\) exhibit similar features. B) Root mean square fluctuation for the angular distribution, \(\theta_m\) for different aspect ratios. The theoretical prediction (shown in red squares with dashed lines) agree  reasonably well for longer rods [see Section \ref{sec:dft}, \eq{fluctheo_prediction}]. C) Comaparison of the angular fluctuation with packing fractions between a  nematic monolayer and 3D bulk systems, for \(A=5.0\).  The light blue dashed vertical lines indicate the phase boundaries for bulk systems. For bulk smectic and crystal phases, the values of \(\eta_{mono}\) (as defined in the text) were used.
    }
    \label{fig:fig4_ornt_dist}
\end{figure*}

It is tempting to seek a comparison with strictly two-dimensional (2D) systems where the celebrated KTHNY theory \cite{kosterlitz1978two,kosterlitz2016kosterlitz,halperin1978theory,nelson1979dislocation,young1979melting} postulates that freezing occurs via an intermediate \textit{hexatic} phase. Our systems, however, are not strictly 2D since the rods can freely rotate in three dimensions. Moreover, the projection of the spherocylinders onto the constraining 2D plane leads to an ensemble of shape-disperse ellipses (with different eccentricities and major and minor axes) residing on a 2D plane. Whether or not the nematic-solid freezing transition of monolayer rods involves a similar intermediate \textit{hexatic} phase remains an unresolved issue as this calls for further large-scale simulations \red{at more closely spaced packing fractions} combined with a systematic finite-size scaling analysis.

Let us now focus on locating the critical packing fractions (\(\eta_c\)) for the nematic-solid freezing transition.
These are obtained from the equation of state (EOS) using standard compression and expansion runs. The results are given in \fig{fig:fig3_eos}. In the presence of a phase transition, the expansion and compression  give rise to different values of the pressure at different densities, which results in a hysteresis in the EOS plots. In our case, it is indicative of the nematic-solid transition being \textit{first-order}.  Interestingly, the hysteresis in the plots for various aspect ratios occurs around the same packing fractions \(\eta \approx 0.82\) (see Fig. \ref{fig:fig3_eos} A,B,C,D) suggesting that the freezing transition is {\em independent} of the rod aspect ratio $A$. This agrees with what is observed in 3D bulk systems of hard (sphero)cylinders \cite{bolhuis_frenkel,mcgrother1996re,lopes2021phase} as well as discotic particles \cite{veerman1992phase,marechal2011phase} where  freezing of a nematic fluid into a positionally ordered phase (smectic, columnar or crystal) happens at a universal packing fraction, independent of the rod aspect ratio.

A shape-independent freezing transition as demonstrated from our simulations is also consistent with a simple theoretical model, as discussed in Section \ref{sec:dft}. \red{An anisotropy-independence  was also established for strongly confined hard ellipses \cite{basurto2020ordering,basurto2021anisotropy}}. 

The transition values  extracted from the EOS are compiled in Table \ref{tab:table_transitionPF}. \red{The critical packing fraction is found to be considerably higher compared to the liquid-solid phase transition observed for 2D hard or soft discs, which occurs around 0.7-0.72 \cite{mak2006large,huerta2006freezing,khali2021two}. This demonstrates that the scenario of flat monolayer of spherocylinders does not conform to a 2D fluid hard-discs as one would naively expect.  We attribute this difference to the role of orientational fluctuations exercized by the rods which are weak but non-zero (the latter would make the orientational entropy per rod diverge). These residual fluctuations are known to have a non-negligible impact on the equation of state of hard-body liquid crystals, even at elevated packing conditions \cite{wensink2004equation}. The effect seems far less prominent for a monolayer of hard ellipsoids where the liquid-solid transition does converge to that of a 2D system of hard disks, at least for small aspect ratio \cite{varga2016effect}. Unlike spherocylinders,  ellipsoids are rounded and tapered which means that small orientational excursions away from the plane normal direction do not alter the  circular cross-section of the ellipsoid projected on the plane. Positional-orientational correlations only emerge when these angular fluctuations $\theta_{m}$ exceed the aspect-ratio of the ellipsoids. In this regime, at larger aspect ratio, the freezing transition was found to drop to lower packing fractions, in contrast to our observation on spherocylinders \cite{varga2016effect}. This points to subtle, non-trivial differences between spherocylinders and ellipsoids which are also known relevant for the thermodynamic stability of smectic liquid crystals \cite{king2023promotes}. 

It is noteworthy that the critical packing fractions at freezing reported in Table I are remarkably  close to the random-close packing fraction ($\eta_{\rm RCP} =0.826 $) reported for randomly jammed systems of hard discs in 2D \cite{atkinson2014existence} although a well-argumented relationship between these two cases can not be established at this stage.

}


\subsection{Orientational fluctuations}


Let us now focus on the orientational fluctuations of the spherocylinders in the different phases. For all aspect ratios (\(A\)) and packing fractions (\(\eta\)) considered in our simulations, the nematic director, \(\hat{n}\), coincides with the plane normal \(\hat{z}\). Orientational fluctuations for each rod are defined by the polar angle $\theta$ between the rod orientational unit vector $\vec{s}$ and $\hat{n}$.  The distribution of the azimuthal angle $\varphi$ remains uniform which demonstrates that, as expected, the orientational order is of uniaxial symmetry.  We define the fluctuation strength via the polar angular distribution via the mean-squared polar angle  \(\langle \theta^{2}  \rangle = \theta_m^2 = \int_{0}^{1} d (\cos \theta)  \theta^{2} P(\cos \theta)/\int_{0}^{1} d (\cos \theta )  P(\cos \theta)\), where the denominator is normalised to \(1\). The distribution $\langle P(\cos \theta) \rangle$ is obtained from binning the dot product $ \vec{s}_{i} \cdot \hat{z} =\cos \theta_{i} $ for each rod $i$ using equidistant bins and taking a thermal average indicated by the brackets. The corresponding orientational distribution in \(\theta\) is shown in \fig{fig:fig4_ornt_dist}A) for a specific aspect ratio of \(A = 5.0\). The distribution is peaked around \(0\degree\) which is also observed for three-dimensional bulk liquid crystal phases based on the same spherocylinder model that we also simulated. The strength of the orientational fluctuations, depicted in   \fig{fig:fig4_ornt_dist}B, decreases monotonically with packing fraction. When scaled with the aspect ratio $A$ the curves collapse onto a mastercurve at high packing, that follows from a density functional theory that we describe in Section \ref{sec:dft}. This suggests that the fluctuations are shape universal and are controlled principally by the rod packing. It is noted from  \fig{fig:fig4_ornt_dist}C that the scaling of the fluctuations with density differs from that of a 3D bulk smectic phases  most likely because of the neglect of out-of-plane as well as (Helfrich-type) planar undulations that are ignored in our monolayer model. This additional fluctuations in positional space are likely to enhance the orientational ones.  One should also bear in mind that the planar density is not the same as the 3D bulk density. In fact, the relationship between the two is $\eta_{mono} = \eta_{3D} \lambda_{Sm}/(L+D)$ with $\lambda_{Sm}$ the smectic spacing which itself depends on the 3D bulk packing fraction. In fact, the values of layer spacing \(\lambda_{Sm}\) decreases marginally with 3D bulk packing fraction \(\eta_{3D}\). In  \fig{fig:fig4_ornt_dist}C, the angular fluctuations for smectic (Sm) and crystal (K) phases are plotted against \(\eta_{mono}\), which indicates that at high enough packing fractions, the fluctuation strength for bulk smectic and crystal phases tends to converge to the values for the monolayer.

\section{from neutral to weakly charged rods} \label{sec:electro}


We will now move on to the case of charged rods and explore the role of electrostatic twist, defined by the effective electrostatic potential \eq{eqn:eqn1_el}.  
As discussed previously, at high packing fraction volume exclusion drives  configurations where the particles align close to the surface normal of the constraining plane.  On the other hand, the electrostatic interaction \(U_{el} \propto 1/|\sin \gamma|\) depends inversely on the relative angle between the particles and diverges when rods are strictly parallel ($\gamma =0$).

In order to reduce the energy of the system, the rods will resist locally parallel configurations in favor of  twisted motifs. Therefore, there is a direct competition between volume exclusion (WCA part) and electrostatic interactions which respectively prefer and disfavors parallel rod alignment. It should be noted, however,  that the simulations for charged rods are technically difficult in view of the divergence of the effective potential for parallel configurations $\gamma =0$. In order to overcome this, we excluded  large interparticle forces below a certain cut-off angle $\gamma_{\rm cut}$. Note that we are using a value of \(\gamma_{\rm cut} = 10^{-5}\), and the small value of \(\gamma_{\rm cut}\) practically allows for the complete electrostatic calculations for all packing fractions at small values of \(u\). Also, to have an quantitative idea of the electrostatic strength, we note that, for \(A=2.0, \eta \approx 0.5, u=0.1, \kappa = 1.0, \langle U_{el}\rangle \approx 0.1\langle U_{WCA}\rangle \), where the angular brackets denote ensemble averages.

\begin{figure}[h!]
    \centering
    \includegraphics[width=0.8\linewidth]{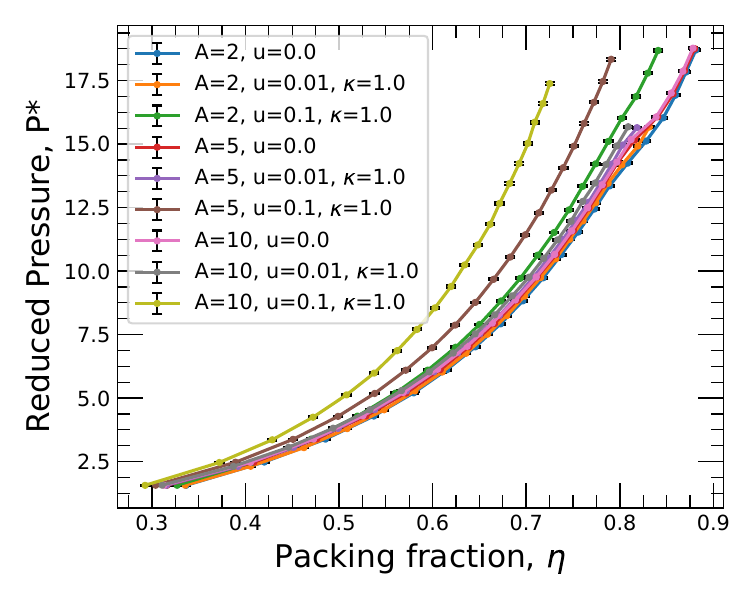}
    \caption{ Equation of state in terms of reduced preesure $P^{\ast}$ versus packing fraction $\eta$ for a rod monolayer for both neutral and charged with various aspect ratios $A$. The electrostatic amplitude is given by $u$ while $\kappa$ denotes the Debye screening length in units of the rod diameter. }
    \label{fig:EOS_Lall_uall_kall}
\end{figure}

As the packing increases and particles align more strongly with one another, the electrostatic potential raises the system energy as well as the pressure compared to the neutral case, as expected (\fig{fig:EOS_Lall_uall_kall}.  Therefore, including electrostatics  leads to a reduction of the isothermal compressibility and an increase effective rod crowding which results in smaller fluctuations. As expected, the reduction of compressibility becomes systematically stronger as the rod aspect ratio $A$ grows larger and double layer overlap becomes  more prominent as evident from strongly charged rods with $u =0.1, A=10$ (Fig. \ref{fig:EOS_Lall_uall_kall}). 

\begin{figure}[]
    \centering
    \includegraphics[width=0.95\linewidth]{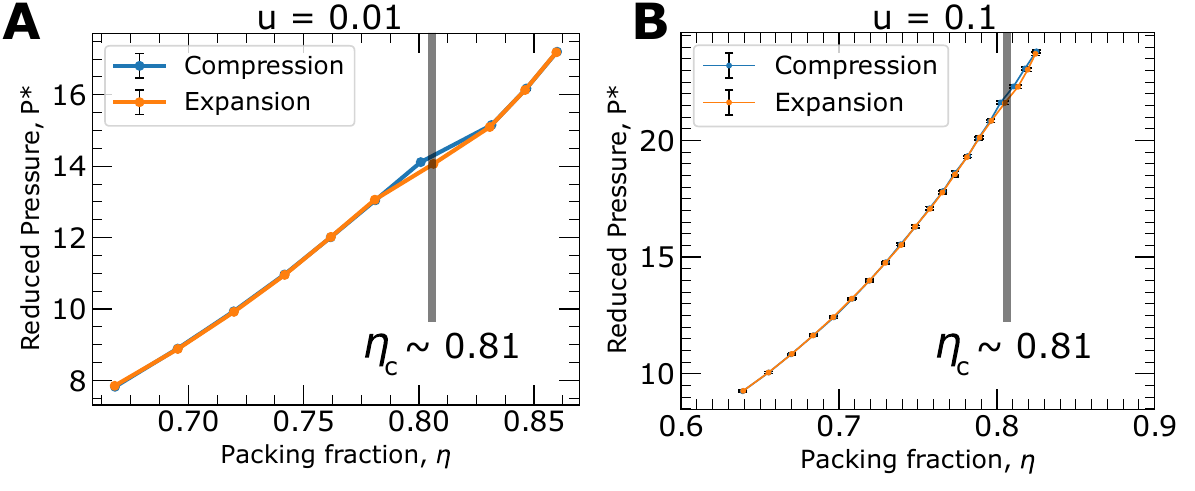}
    \caption{Nematic-solid freezing transition for charged rods with \(A=2.0\). Shown are the equations-of-state from both compression and expansion runs for weakly and strongly charged rods, \(u=0.01\) and \(u=0.1\) respectively.  }
    \label{fig:effect_of_el_on_etac}
\end{figure}



Next, we take a close look at the effect of the electrostatic repulsion on the freezing transition, structure and symmetry of the solid phase. In \fig{fig:effect_of_el_on_etac}A) and \ref{fig:effect_of_el_on_etac}B) we show the EOS from compression and expansion scheme as discussed earlier, using two different strengths of electrostatic interaction \(u=0.01\) and \(u=0.1\). Both cases feature a discontinuity around the same packing fraction \(\eta_c \approx 0.81\) which suggests the freezing transition of a monolayer rod fluid is little affected by the presence of angle-dependent electrostatic repulsive forces, at least for the regime of amplitudes explored here ($u \leq 0.1$). 

\begin{figure}[]
    \centering
    \includegraphics[width=0.95\linewidth]{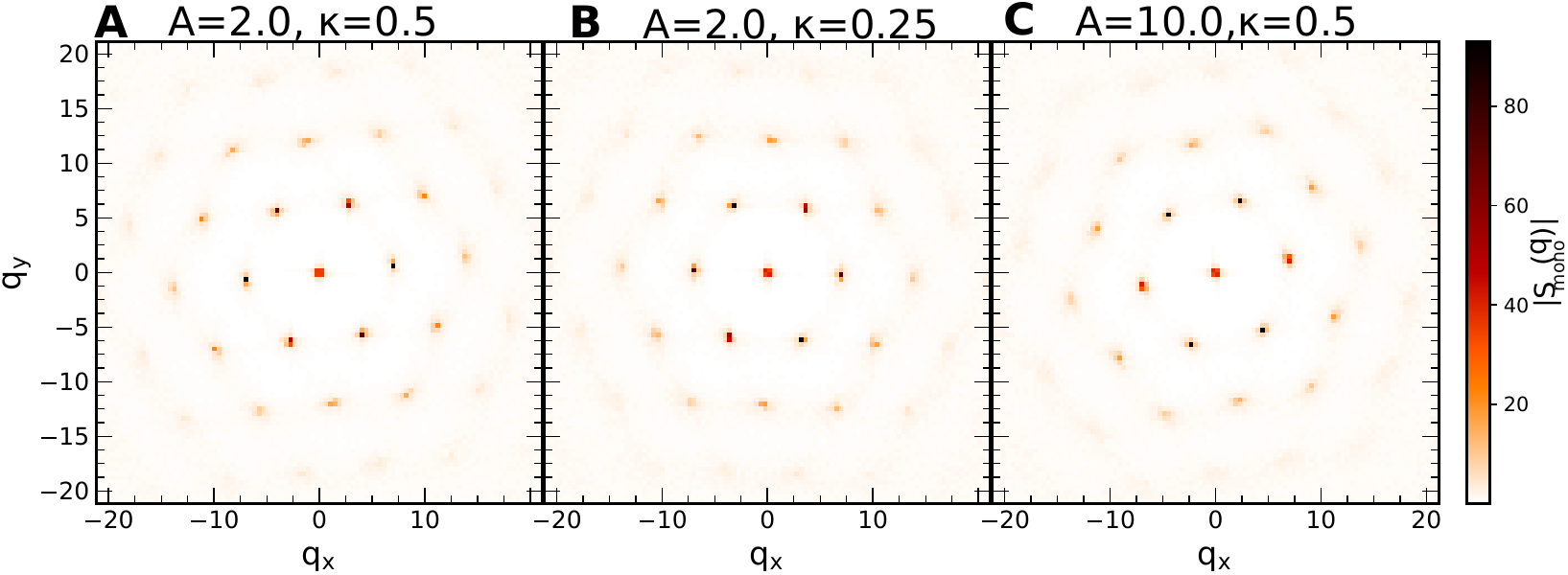}
    \caption{ Structure factor of a monolayer solid phase of charged rods at fixed packing fraction \(\eta \approx 0.83, u=0.01\). Shown are results for different aspect ratio $A$ and Debye screening parameter $\kappa$.  A hexagonal crystal arrangement is present for  all cases. }
    \label{fig:Str_fac_combined}
\end{figure}

The symmetry and degree of positional ordering  of the system can  be quantified from the static structure factor $S_{\rm mono}(\vec{q})$ defined as follows,
\begin{equation}
    S_{\rm mono}(\vec{q}) = \frac{1}{N} \left \langle \sum_{i=1}^N\sum_{j=1}^{N} e^{-i\vec{q} \cdot \left(\vec{r}_i -\vec{r}_j\right) } \right \rangle
    \label{smono}
\end{equation}
where \(\langle...\rangle\) denotes a thermal average and $\vec{q}$ the wavevector.

The structure factors are depicted in  \fig{fig:Str_fac_combined}A),B) and C) and the appearance of a hexagonal arrangement of Bragg peaks  confirming the hexagonal lattice structure we already inferred in real space from \fig{fig:Fig1_initial}(D). 
In reciprocal space, the  Bragg peaks are separated at a distance  \(|q| \approx 6.98\) which corresponds to a real-space lattice distance \(a_0 = 2\pi/q = 0.9\), consistent with the first peak of the radial distribution function plot for the neutral case (Fig. \ref{fig:Fig2_gofr}). The results confirm that the effect of electrostatic repulsion and changes of the ionic strength through the screening parameter $\kappa$ do not lead to qualitative changes in the microstructure of the solid phase.  We do point out that the notion that structure formation is dominated by steep-repulsive (WCA) interactions with the electrostatics acting as a mere perturbation only holds for the weak electrostatic repulsion regime where the line potential Eq.\ref{eqn:eqn1_el} holds validity. As such we can not make any claim as to what happens for crowded monolayer systems of strongly charged rods at weak electrostatic screening since this necessitates a more sophisticated treatment of the problem of electrostatics at charged colloidal surfaces \cite{andelman2006introduction,agra2004interplay}.



\section{Simplified DFT for freezing  of a monolayer rod fluid } \label{sec:dft}

In order to complement our computer simulations with a theoretical basis and provide an  argument for the shape universality of the freezing transition, we wish to explore density functional theory (DFT) which is a reference theoretical framework to address freezing in classical fluids \cite{evans1979nature}. Within the context of liquid crystals, however,  the formal theory  is technically cumbersome as it requires knowledge of the direct correlation function for anisotropic particles which depends on multiple angles \cite{workman1973integral}. Our approach will be much simpler as we focus on well-established volume exclusion arguments inspired by Onsager's theory for nematic fluids \cite{onsager1949effects,mulder1987density,xu1992nematic,allen1993hard}.

The key quantity of interest is the one-body density which is the ensemble-average of the microscopic  density $\rho( \bor ; \oma)$ of rods residing in a smectic plane parameterized by position $\bor$ and orientation $\oma$. As evident from our simulations, the monolayer nematic transitions into a solid with long-ranged in-plane positional order. At the onset of freezing transition the one-body density can be expanded around  reference solution for the fluid state  in terms of some density modulation with infinitesimally small amplitude $\epsilon \ll 1$, 
\beq
\rho( \bor ; \oma) = \rho \left [ P(\oma) + \epsilon P_{q} (\oma) e^{i {\bf q} \cdot \bor} \right ].
\eeq
with $\rho$ denoting the overall rod concentration within the plane and $P_{q}(\oma)$ some angular probability which  may be affected  by the wavevector ${\bf q}$ of the in-plane density modulation. In working out the DFT we closely follow the approach of Ref. \cite{wensink2023elastic} and refer the reader to that work for details. The basic assumption in the theory is that rods are represented as purely hard cylinders which interact through pair collisions alone with an ad-hoc density rescaling accounting for higher virial terms using Lee-Parsons theory. 

It is generally safe to assume that the orientations are not affected too much by the development of positional order. This is also confirmed in our simulation by the fact that near the freezing transition, the values of the orientational order (\(S\)) are almost same (see Fig. \ref{fig:Fig3_nemop}, near the shaded region). Then we may localize the onset of freezing from  a divergence of the static structure factor $S_{\rm mono}(q)$ signalling loss of local stability of the uniform fluid, 
\begin{align}
& S_{\rm mono}^{-1}( q ) =   1 + 8\eta g(\eta) \left \{ \frac{J_{1} (q) }{q}  + \langle \langle \mathcal{F} ({\bf q}; 
\oma) \rangle \rangle_{P}  \right \}  =0.
\label{sinfrozen}
\end{align}
with $J_{1}(x)$ a Bessel function of the first kind and $g(\eta)= (1 - \tfrac{3}{4} \eta)/(1- \eta)^{2}$ a Lee-Parsons renormalization factor. The first term between brackets stems from the volume exclusion between two strictly parallel cylinders. The second contribution accounts for the orientational fluctuations of the rods and features an  coupling between the density wave and the azimuthal rod angle  \cite{wensink2023elastic}. The brackets denote a double angular average in terms of some angular probability  which for reasons of tractability we choose to express as a delta distribution in the polar angle $\theta$,
\beq
P(\oma ) \sim  \frac{\delta (\theta - \theta_{m})}{\sin \theta_{m}} \frac{1}{2 \pi}.
\label{ffact}
\eeq
with $\theta_{m} = \langle \theta^{2}\rangle^{1/2}$ the mean squared amplitude of the polar angular fluctuations assumed Gaussian \cite{wensink2023elastic}. The azimuthal angle $\varphi$ around layer normal ${\bf \hat{z}}$ is supposed to remain uniform for weakly developed positional order and  the fluctuation terms reduces to, 
\begin{align}
& \langle \langle \mathcal{F} ({\bf q}; \oma) \rangle \rangle_{P}  \sim  \frac{4}{\pi^{2}} (A \theta_{m}) \int_{0}^{2 \pi} \frac{d\varphi}{2 \pi} j_{0} [ \tfrac{q }{2} (A \theta_{m}) \cos \varphi  ]. 
\label{fsimple}
\end{align}
 At thermodynamic equilibrium, the mean squared angular fluctuations $\theta_{m}$ is connected to the in-plane rod packing fraction  $\eta$ via  \cite{wensink2023elastic}
\beq 
A \theta_{m} \sim \left ( \frac{\pi}{2} \right )^{1/2} \frac{1}{\eta g(\eta)},
\label{fluctheo_prediction}
\eeq
This prediction agrees quite well with our simulation results, as demonstrated in
\fig{fig:fig4_ornt_dist}B at least for long rods. The theoretical model deviates from the simulation results for very short rods \(A = 2\)
which are poor nematogens that do not form LC phases in bulk. For short spherocylinders, pronounced endcap correlations becomes prominent that are not accurately described by
the simplified excluded volume term featuring in \eq{sinfrozen}. The static structure factor \eq{sinfrozen} is {\em independent} from the rod aspect ratio $A \gg 1$ which implies that the freezing transition of the monolayer rod fluid is unaffected by the rod shape which is in agreement with our simulations (Table I). The case of perfectly parallel hard cylinders has been  analyzed by Mulder \cite{mulder1987density} and can be reproduced  from  setting the mean squared fluctuations to zero ($\theta_{m} =0$). Solving \eq{sinfrozen} entails finding the lowest (real-positive) packing fraction $\eta_{c}$ with the associated wavenumber indicated by $q_{c}$
gives the following solution $\eta_{c} = 0.58$ ($q_{c} = 5.13$) which clearly falls below the values from Table \ref{tab:table_transitionPF}. Including the orientational fluctuations encapsulated by a non-zero $\langle \langle \mathcal{F} \rangle \rangle_{P}$  we find a much more accurate prediction $\eta_{c} = 0.81$  ($q_{c} = 0.513$) which is in quantitative agreement with the values reported in Table \ref{tab:table_transitionPF}. 

\red{The static structure factor for the fluid range is easily computed from \eq{sinfrozen} and is compared with simulation results in \fig{fig:Fig2_gofr}. We see that the qualitative trends are very similar but the peak position is shifted somewhat because of the soft-core model used in simulation which leads to the rod cores overlapping. }

\section{Conclusion} \label{sec:conclusion}

Inspired by recent experimental studies of rod-shaped colloids residing on flat substrates we have explored self-assembly of freely rotating  soft spherocylinders with mass centers confined to a flat plane thus resembling a single membrane within a stacked smectic or lamellar architecture. Upon increasing the rod packing fraction within the plane,  we identify two principal states, i) a nematic fluid at low to medium densities  where the system possesses global orientational ordering but no long-ranged positional ordering and ii) a crystalline solid phase at elevated particle packing characterized by strong orientational and hexagonal positional ordering. 

We find that the critical packing fraction for the nematic-solid transition  \(\eta_c\ \approx 0.82\) is independent of the aspect ratio of the spherocylinders suggesting freezing to be shape universal. \red{Furthermore, we conclude that freezing of a monolayer fluid of spherocylinders occurs at a considerably higher packing fraction compared to  more commonly studied previous cases such as  2D hard disks and hard ellipsoids in planar confinement}.   A simple density functional theory  provides a theoretical underpinning for the shape universality of the freezing transition. 
As far as the orientational order is concerned our simulations demonstrate that the strength of the orientational  scales differently with density than in 3D bulk smectics. More specifically, the reduction of the orientational fluctuations with increased packing is much weaker for monolayer rod fluids than for 3D bulk smectic phases  \cite{bolhuis_frenkel,mcgrother1996re,lopes2021phase}.   The overall phenomenology is robust against the inclusion of electrostatic forces, at least up to small electrostatic amplitudes. The presence of electrostatic twist imparted by overlapping electric double layers of adjacent rods when they align does not alter the crystal morphology at high in-plane packing fraction. We conclude that electrostatic repulsion merely leads to a reduced compressibility and enhance rod crowded which boosts rod alignment without affecting the critical packing fraction at freezing.

Our results contribute to the fundamental understanding of the stability of complex smectic phases and their subtle in-plane (orientational) microstructure. Our work also has potential technological and experimental relevance, for instance, to understand the phase behaviour of gold nanorods on 2D substrates or polymer-dispersed liquid crystal monolayers.  Future avenues of research could relate to unveiling the nature of phase transition, in particular the existence of a hexatic phase intervening between the fluid and solid states \cite{berezinskii1971zh,nelson1979dislocation}, and to highlighting the effect of molecular chirality \cite{chattopadhyay2024stability} and rod backbone flexibility which plays a role for certain stiff biopolymers such as cellulose nanofibers \cite{eichhorn2010current}, amyloid fibrils \cite{bagnani2018amyloid} and monodisperse filamentous virus rods. The latter are known to form SmA phases \cite{dogic1997smectic} whose elastic and microstructural properties seem to largely follow a single fluid layer description \cite{wensink2023elastic}. In line with previous simulations projects on active matter involving some of us another  future aplication of our model could be studying the steady state behaviour of active and passive \cite{chattopadhyay2023two} polymer mixtures responding to planar-type confinement \cite{venkatareddy2023effect}.

\begin{acknowledgments}

JM acknowledges the Council of Scientific and Industrial Research (CSIR), India for a fellowship and is grateful to Dharanish Rajendra and Parvathi K for insightful discussions. PKM thanks the Department Of Science \& Technology (DST), India for financial support and the Science and Engineering Research Board (SERB) for funding and computational support.
\end{acknowledgments}

\section*{Conflict of Interest}
The authors have no conflicts to declare.

\section*{Data Availaiblity}
 The data that support the findings of this study are available from the corresponding author upon reasonable request.



\bibliography{ref}


\appendix

\renewcommand\thefigure{\thesection.\arabic{figure}}    
\
%
\end{document}


\newgeometry{top=1.50 cm,bottom=1.50 cm,left=1.50 cm,right=1.25 cm}

\maketitle

\section{Compressibility reduction for charged rods}

As expected, the compressibility is reduced for a system of charged rods for small electrostatic interaction amplitudes that we explored ($u<0.1$). This is shown in \fig{fig:EOS_Lall_uall_kall} below. The effect is most clearly reflected for long rods $A=10$.

\begin{figure}[h!]
    \centering
    \includegraphics[width=0.6\linewidth]{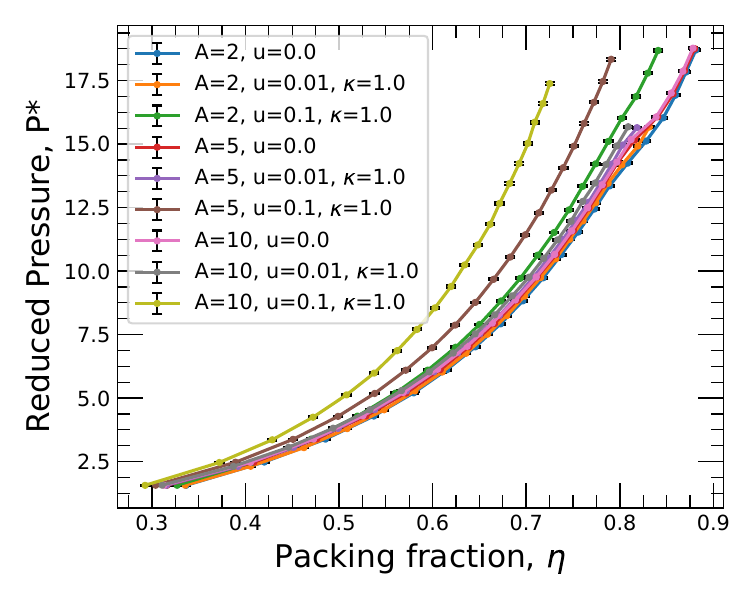}
    \caption{ Overview of the equation of state in terms of reduced preesure $P^{\ast}$ versus packing fraction $\eta$ for a rod monolayer for both neutral and charged with various aspect ratios $A$. The electrostatic amplitude is given by $u$ while $\kappa$ denotes the Debye screening length in units of the rod diameter. }
    \label{fig:EOS_Lall_uall_kall}
\end{figure}



\section{Conically degenerate tilt angle for charged rods}

The orientational distribution and the tilt angle for  charged rods  show qualitatively similar behaviour to the neutral case ($u=0$). For the weak amplitudes considered ($u\leq 0.1$) the electrostatic interaction can be considered as a perturbation for the considered set of values of \(u\). In \fig{fig:OD_compare_tilt_angle_el_on_off_compare} we demonstrate that the tilt angle \(\theta_t\) decreases with the electrostatic interaction strength \(u\) for a specific value of \(A\) and \(\eta\).

\begin{figure}[h!]
    \centering
    \includegraphics[width=0.6\linewidth]{Figures/OD_compare_tilt_angle_el_on_off_compare.pdf}
    \caption{(A)-(C) Effect of twist electrostatic repulsion on the rod orientational probability for a variety of rod aspect ratios \(A=2.0,5.0\) and \(10.0\), electrostatic amplitudes ranging from neutral to weakly charged rods (\(u=0.0,0.01,0.1\)) and Debye screening lengths $\kappa$. (D) Variation of the tilt angle  with packing fraction $\eta$.  }
    \label{fig:OD_compare_tilt_angle_el_on_off_compare}
\end{figure}